\documentclass[aps,groupedaddress,showpacs,floatfix]{revtex4}
\usepackage{amssymb}
\usepackage{amsmath}
\usepackage{graphicx}
\begin{document}

\thispagestyle{empty}

\title{Scaling properties of (2+1) directed polymers in the low temperature limit}

\author{Victor Dotsenko}

\affiliation{LPTMC, Sorbonne Universit\'e, Paris, France}

\date{\today}

\begin{abstract}
In  terms of the replica method we consider the low temperature limit of (2+1) directed polymers 
in a random potential. The proposed approach allows to compute  
the scaling exponent $\theta$ of the free energy fluctuations as well as the 
left tail of its probability distribution function. It is argued that 
$\theta = 1/4$ which is slightly different from the zero-temperature numerical
value which is close to $0.241$.
\end{abstract}


\maketitle

\section{Introduction}

In contrast to the achievements in the studies of one-dimensional directed polymers
as well as the other systems belonging to the so called KPZ universality class
\cite{KPZ,Corwin,Borodin,Rev,Takeuchi}, not so much is achieved in understanding of the scaling properties 
of the so called (2+1) model of directed polymers which describes the fluctuations 
of an elastic string directed along the time axes and which passes through 
a random medium in the three-dimensional space
\cite{(2+1)-2009,(2+1)-2017a,(2+1)-2017b,(2+1)-2019,(2+1)-2020,(2+1)-exper-2015}.
Extensive numerical simulations as well as experimental studies rather convincingly demonstrate that in the 
limit of large times $t$ the free energy fluctuations  
of such polymers scale as $ t^{\theta}$ with the scaling exponent 
$\theta \simeq 0.241$ which is very close but not equal to $1/4$
 \cite{numerics1,numerics2,numerics3}. On the other hand,
in the recent analytical study \cite{(2+1)-highT} it has been argued that 
in the {\it high-temperature limit} this scaling exponent is equal to $\theta = 0.5$.
If this statement is correct, it would mean that the scaling exponent of the (2+1) system
is {\it not universal} being temperature dependent in the drastic difference with the 
corresponding (1+1) model where this exponent is universal and equal to $1/3$ at all 
temperatures.

In this paper  in terms of the replica technique we study the scaling properties
of the two-dimensional directed polymers in the low-temperature limit. 
The idea is to estimate the replica partition function $Z(N,t)$ 
(which is the disorder average of the 
$N$-th power of the partition function) in the limit of $N \gg 1$ which makes possible  
to reconstruct the {\it left tail} asymptotic
of the free energy distribution function $P(F\to -\infty)$. Assuming that this 
(unknown) distribution function is defined by the only energy scale
we find that the time scaling exponent of the free  energy fluctuations is  $\theta=1/4$. 
We see that this value is somewhat  different from the 
zero temperature numerical results. To explain this discrepancy I would suggest that 
in the numerical simulations the exact value of the scaling exponent $\theta = 1/4$ 
is distorted by the presence of a logarithmic prefactor which is common situation 
for the two-dimensional systems.

In Section II we define the  model and describe the general ideas
of the considered replica approach. 
In Section III we present the main lines of the "energy optimization method"  
used for estimation of the replica partition function in the large $N$ limit.
In Section IV we  demonstrate how this method works in the case of (1+1) directed polymers
and show that here it perfectly reproduces well known exact results.
In section V we concentrate on (2+1) directed polymer model for which up to now 
no exact results are known. Here we demonstrate that in the low temperature limit 
and at $N \gg 1$ the replica partition function 
$Z(N,t) \sim \exp\bigl\{ u \, (\beta \, \epsilon)^{-2}  \, (\beta N)^{4} \, t\bigr\}$,
where $\beta$ is the inverse temperature, $u$ is the strength of the random potential
and $\epsilon$ is its spatial correlation length (see eqs.(\ref{3})-(\ref{4})). 
Correspondingly, 
for the left tail of the free energy distribution function
one obtains
\begin{equation}
\label{1}
P(F\to -\infty) \; \sim \; 
\exp\biggl\{- \biggl(\frac{|F|}{\bigl(\sqrt{u}/\beta \epsilon\bigr)^{1/2}  \; t^{1/4}}\biggr)^{4/3}\biggr\}
\end{equation}
so that the free energy fluctuations scales as $|F| \, \sim \, t^{1/4}$.
Finally, Section VI is devoted to the discussion of the obtained results.

\section{The model and the replicas approach} 

We consider the model of directed polymers defined in terms
of an elastic string described by the two-dimensional vector
$\boldsymbol{\phi}(\tau) \equiv \bigl(\phi_{x}(\tau), \, \phi_{y}(\tau)\bigr)$ 
directed along the $\tau$-axes within an interval $[0,t]$ 
which passes through a random medium
described by a random potential $V(\boldsymbol{\phi},\tau)$. 
The energy of a given polymer's trajectory
$\boldsymbol{\phi}(\tau)$ is
\begin{equation}
\label{2}
H[\boldsymbol{\phi}(\tau); V] = \int_{0}^{t} d\tau
\biggl[\frac{1}{2} \Bigl(\partial_\tau \boldsymbol{\phi}(\tau)\Bigr)^2
+ V\bigl(\boldsymbol{\phi}(\tau),\tau\bigr)\biggr];
\end{equation}
Here the disorder potential $V[\boldsymbol{\phi},\tau]$
is supposed to be Gaussian distributed with a zero mean
$\overline{V(\boldsymbol{\phi},\tau)}=0$
and the correlation function
\begin{equation}
\label{3}
\overline{V(\boldsymbol{\phi},\tau)V(\boldsymbol{\phi}',\tau')} = 
u \, \delta(\tau-\tau') \delta^{(2d)}_{\epsilon}(\boldsymbol{\phi}-\boldsymbol{\phi}')
\end{equation}
The parameter $u$ is the strength of the disorder and $\delta^{(2d)}_{\epsilon}(\boldsymbol{\phi})$ 
is the two-dimensional "finite-size $\delta$-function":
\begin{equation}
\label{4}
\delta^{(2d)}_{\epsilon}(\boldsymbol{\phi}) \; = \; \left\{ 
\begin{array}{ll}
\frac{1}{\pi \epsilon^{2}} 
\; ,  
\; \; \; \;
\mbox{for} \; |\boldsymbol{\phi}| \, \leq \, \epsilon 
\\
\\                          
0 \; , 
\; \; \; \; \; \; \; 
\mbox{for} \;  |\boldsymbol{\phi}| \, > \, \epsilon                 
\end{array}
\right.
\end{equation}
We do not introduce here "true" (zero-size) two-dimensional $\delta$-function for the reasons which will be 
explained later. 


\vspace{5mm}

For a given realization of the random potential $ V(\boldsymbol{\phi},\tau)$
the partition function of the considered system (with fixed boundary conditions) is
\begin{equation}
\label{5}
Z({\bf r}, t) = \int_{\boldsymbol{\phi}(0)=\bf{0}}^{\boldsymbol{\phi}(t)={\bf r}} 
{\cal D}\boldsymbol{\phi}(\tau)
\exp\bigl\{-\beta H[\boldsymbol{\phi}(\tau), V]\bigr\} \; = \; \exp\bigl\{-\beta F({\bf r}, t)\bigr\}
\end{equation}
where $\beta$ is the inverse temperature,
$F({\bf r},t)$ is the free energy which is a random quantity 
and the integration is taken over all
trajectories $\boldsymbol{\phi}(\tau)$
starting at $\bf{0}$ (at $\tau = 0$) and ending at a point ${\bf r}$ 
(at $\tau = t$). 

For simplicity, in what follows we are going to consider the problem with
the zero boundary conditions: $\boldsymbol{\phi}(\tau=0)=\boldsymbol{\phi}(\tau=t)={\bf 0}$.
The free energy probability distribution function
$P(F)$ of this system can be studied in terms of the integer moments 
of the above partition function, eq.(\ref{5}):
\begin{equation}
\label{6}
\overline{Z^{N}} \equiv Z(N,t) \; = \; 
\prod_{a=1}^{N}\int_{\boldsymbol{\phi}_{a}(0)=0}^{\boldsymbol{\phi}_{a}(t)=0}
{\cal D}\boldsymbol{\phi}_{a}(\tau) \;
\overline{
	\Biggl(
	\exp\Bigl\{-\beta \sum_{a=1}^{N} H[\boldsymbol{\phi}_{a}(\tau), V]\Bigr\}
	\Biggr)}
\; = \;
\int_{-\infty}^{+\infty} dF \, P(F) \, \exp\bigl\{-\beta N F\bigr\}
\end{equation}
where $\overline{(...)}$ denotes the averaging over 
the random potentials $V[\boldsymbol{\phi},\tau]$ 
Performing this simple Gaussian averaging we get
\begin{equation}
\label{7}
Z(N,t) \; = \; \prod_{a=1}^{N}\int_{\boldsymbol{\phi}_{a}(0)=0}^{\boldsymbol{\phi}_{a}(t)=0}
{\cal D}\boldsymbol{\phi}_{a}(\tau) \;
\exp\Bigl\{-\beta  
H_{N}[\boldsymbol{\phi}_{1}(\tau), \, \boldsymbol{\phi}_{2}(\tau), \, ... \, , \boldsymbol{\phi}_{N}(\tau)]
\Bigr\}
\end{equation}
where
\begin{equation}
\label{8}
\beta H_{N}[\boldsymbol{\phi}_{1}(\tau), \, \boldsymbol{\phi}_{2}(\tau), \, ... \, , \boldsymbol{\phi}_{N}(\tau)] =  \int_{0}^{t} d\tau
\Biggl[
\frac{1}{2} \beta \sum_{a=1}^{N} \Bigl(\partial_\tau \boldsymbol{\phi}_{a}(\tau)\Bigr)^2
\; - \;
\frac{1}{2} \beta^{2} \, u \,
\sum_{a,b=1}^{N}\delta^{(2d)}_{\epsilon}\bigl(\boldsymbol{\phi}_{a}(\tau) - \boldsymbol{\phi}_{b}(\tau)\bigr)
\Biggr];
\end{equation}
is the replica Hamiltonian which describes $N$ elastic strings
$\bigl\{\boldsymbol{\phi}_{1}(\tau), \, \boldsymbol{\phi}_{2}(\tau), \, ... \, , \boldsymbol{\phi}_{N}(\tau)\bigr\}$
with the attractive interactions 
$\delta^{(2d)}_{\epsilon}\bigl(\boldsymbol{\phi}_{a} - \boldsymbol{\phi}_{b}\bigr)$, eq.(\ref{4}).
To compute the replica partition function $Z(N,t)$, eqs.(\ref{7})-(\ref{8}), one introduces
the function:
\begin{equation}
\label{9}
\Psi({\bf r}_{1}, \, {\bf r}_{2}, \, ... \, {\bf r}_{N}; \; t) \; = \;
\prod_{a=1}^{N}\int_{\boldsymbol{\phi}_{a}(0)=\bf{0}}^{\boldsymbol{\phi}_{a}(t)={\bf r}_{a}}
{\cal D}\boldsymbol{\phi}_{a}(\tau) \;
\exp\Bigl\{-\beta  H_{N}[\boldsymbol{\phi}_{1}(\tau), \, \boldsymbol{\phi}_{2}(\tau), \, ... \, , \boldsymbol{\phi}_{N}(\tau)]\Bigr\}
\end{equation}
such that 
\begin{equation}
\label{10}
Z(N,t) \, = \, \Psi({\bf r}_{1}, \, {\bf r}_{2}, \, ... \, {\bf r}_{N}; \, t)\big|_{{\bf r}_{a}=0}
\end{equation}
One can easily show that $\Psi({\bf r}_{1}, \, {\bf r}_{2}, \, ... \, {\bf r}_{N}; \, t)$ 
is the wave function of N quantum bosons defined by
the imaginary time Schr\"odinger equation
\begin{equation}
\label{11}
\beta \frac{\partial}{\partial t} 
\Psi({\bf r}_{1}, \, {\bf r}_{2}, \, ... \, {\bf r}_{N}; \; t) \; = \;
\frac{1}{2}\sum_{a=1}^{N} \, \Delta_{a} 
\Psi({\bf r}_{1}, \, {\bf r}_{2}, \, ... \, {\bf r}_{N}; \; t)
\; + \; \frac{1}{2} \, \beta^{3} u \, \sum_{a,b=1}^{N} \delta^{(2d)}_{\epsilon}({\bf r}_{a} - {\bf r}_{b}) \,
\Psi({\bf r}_{1}, \, {\bf r}_{2}, \, ... \, {\bf r}_{N}; \; t)
\end{equation}
where $\Delta_{a}$ is the two-dimensional Laplacian 
with respect to the coordinate ${\bf r}_{a}$.  The corresponding eigenvalue equation 
for the eigenfunctions $\psi({\bf r}_{1}, \, {\bf r}_{2}, \, ... \, {\bf r}_{N})$, 
defined by the relation
\begin{equation}
\label{12}
\Psi({\bf r}_{1}, \, {\bf r}_{2}, \, ... \, {\bf r}_{N}; \; t) \; = \;
\psi({\bf r}_{1}, \, {\bf r}_{2}, \, ... \, {\bf r}_{N}) \, 
\exp\bigl\{-t \, E_{N}\bigr\}
\end{equation}
reads:
\begin{equation}
\label{13}
-2 \beta \, E_{N} \, 
\psi({\bf r}_{1}, \, {\bf r}_{2}, \, ... \, {\bf r}_{N}) \; = \;
\sum_{a=1}^{N} \, \Delta_{a} 
\psi({\bf r}_{1}, \, {\bf r}_{2}, \, ... \, {\bf r}_{N})
\; + \; \kappa \, \sum_{a,b=1}^{N} \delta^{(2d)}_{\epsilon}({\bf r}_{a} - {\bf r}_{b}) \,
\psi({\bf r}_{1}, \, {\bf r}_{2}, \, ... \, {\bf r}_{N})
\end{equation}
where $\kappa = \beta^{3} u$.

\vspace{5mm}

It is at this stage that we are facing the crucial difference of the considered problem
with the corresponding (1+1) one (see Section III). The general solution
of the one-dimensional counterpart of eq.(\ref{13}) is given by the Bethe ansatz 
wave function \cite{Lieb-Liniger,McGuire,Yang}
which is valid only for $U(x) \, = \, \delta(x)$ and which is based on the 
exact two-particle wave functions ($N=2$) solution 
exhibiting  finite value  energy $E_{N=2}$. It is this fundamental property
of (1+1) problem which eventually allows to derive the Tracy-Widom distribution
for the free energy fluctuation
\cite{KPZ-TW1a,KPZ-TW1b,KPZ-TW1c,KPZ-TW2,LeDoussal-2010,VD-TW-2010-1,VD-TW-2010-2}.
The situation in the (2+1) case, eq.(\ref{13}), is much more complicated. 
First of all, in two dimensions there exists no {\it finite} two-particle solution
for the true $\delta$-function potential 
$\lim_{\epsilon\to 0} \delta^{(2d)}_{\epsilon}({\bf r})$, eq(\ref{4}). 
Indeed, for $N=2$ eq.(\ref{13}) (where for simplicity we take $\beta =1$ and $u = \pi$) reduces to
\begin{equation}
\label{14}
\psi''(r) \, + \, \frac{1}{r} \psi'(r)  \, + \, 
\bigl(\pi \delta^{(2d)}_{\epsilon}({\bf r}) \, + \, 2 E_{2}\bigr) \,\psi(r) \; = \; 0
\end{equation}
where  $r = |{\bf r}_{1} - {\bf r}_{2}|$.
It can be easily shown that at $\epsilon \ll 1$ the ground state solution of this equation 
(with $E_{2} < 0$) is 
\begin{equation}
\label{15}
\psi(r) \; \simeq \;  
\left\{ 
\begin{array}{ll}
\psi_{0} \, J_{0}\bigl(\sqrt{1 - 2\epsilon^{2} |E_{2}|} \; r/\epsilon\bigr) 
\; ,  
\; \; \; \; 
\mbox{for} \; r \, \leq \, \epsilon 
\\
\\                          
\psi_{1} \, K_{0}\bigl(\sqrt{2\epsilon^{2} |E_{2}|} \; r/\epsilon\bigr)  \; , 
\; \; \; \; \; \; \; \; \; 
\mbox{for} \;  r \, > \, \epsilon                 
\end{array}
\right.
\end{equation}
where $J_{0}(x)$ and $K_{0}(x)$ are Bessel and Macdonald functions correspondingly,  and
(irrelevant) constants $\psi_{0}$ and $\psi_{1}$ are related by the condition
$ \psi_{0} \, J_{0}\bigl(\sqrt{1 - 2\epsilon^{2} |E_{2}|}\bigr) \; = \; 
  \psi_{1} \, K_{0}\bigl(\sqrt{2\epsilon^{2} |E_{2}|}\bigr)$. The ground state energy 
of this two-particle state is
\begin{equation}
\label{16}
E_{2}(\epsilon) \; \simeq \; 
- \frac{1}{2\epsilon^{2}} \, x_{*}
\end{equation}  
where the number $x_{*} \simeq 0.04$ is the solution of the equation
\begin{equation}
\label{17}
\frac{ J_{0}'\bigl(\sqrt{1 - x}\bigr) \; \sqrt{1 - x}}{J_{0}\bigl(\sqrt{1 - x}\bigr)} \; = \; 
\frac{ K_{0}'\bigl(\sqrt{x}\bigr) \; \sqrt{x}}{K_{0}\bigl(\sqrt{x}\bigr)}
\end{equation}
We see that in the limit $\epsilon \to 0$ (where $\delta^{(2d)}_{\epsilon}({\bf r})$
turns into the "true" $\delta$-function) the ground state energy of this state 
$E_{2}(\epsilon\to 0) \, \to \, -\infty$. 
This simple exercise clearly demonstrate that first, consideration of the 
$N$-particle problem, eq.(\ref{13}) with the true (zero-size) two-dimensional 
$\delta$-function  
makes no sense and second, the ground state energy of this $N$-particle problem must
explicitly depend on the (non-zero) size $\epsilon$ 
of the interaction potential $\delta^{(2d)}_{\epsilon}({\bf r})$, eq.(\ref{4}).

\subsection{Free energy probability distribution function in the replica approach}

Free energy probability distribution function (PDF) $P(F)$ and the replica partition function 
are related according to eq.(\ref{6}):
\begin{equation}
\label{18}
\int_{-\infty}^{+\infty} dF \, P(F) \, \exp\bigl\{-\beta N F\bigr\}
\; = \; Z(N,t)
\end{equation}
The problem is that to recover  $P(F)$ from the above equation (e.g. by inverse Laplace
transform) one has to derive the exact solution for the replica partition function 
$Z(N,t)$ which is hardly possible for the model under consideration.
However if we do not pretend to obtain an entire PDF $P(F)$, but we are interested to  
know only its {\it left tail}, $P(F\to -\infty)$, then the situation somewhat simplifies.
The point is that the asymptotics $P(F\to -\infty)$ is defined by the behavior 
of the replica partition function $Z(N,t)$ in the limit $N\to \infty$ only.

For example, let us suppose that in the limit $N\to \infty$ the replica partition
function, eqs.(\ref{10}), (\ref{12}) has the form
\begin{equation}
\label{19}
Z(N,t) \; \sim \; \exp\Bigl\{g \, (\beta N)^{k} \, t\Bigr\}
\end{equation}
where the factor $g$ depends on the parameters of the model. 
In this case it can be easily shown that the left tail of the function
$P(F)$ has stretched-exponential form,
\begin{equation}
\label{20}
P(F\to -\infty) \; \sim \; \exp\Bigl\{ -\gamma \, |F|^{\alpha}\Bigr\}
\end{equation} 
where it is supposed that $\alpha > 1$. Indeed,
substituting eqs.(\ref{19}) and (\ref{20}) into eq.(\ref{18}) we get
\begin{equation}
\label{21}
\int_{-\infty}^{0} dF \, P(F) \, \exp\bigl\{\beta N |F|\bigr\}
\; + \; 
\int_{0}^{+\infty} dF \, P(F) \, \exp\bigl\{-\beta N F\bigr\}
\; \sim \; 
\exp\Bigl\{g \, (\beta N)^{k} \, t\Bigr\}
\end{equation}
It is clear that due to the factor $\exp\bigl\{-\beta N F\bigr\}$ the second 
integral in the l.h.s of the above relation vanishes in the limit $N\to +\infty$,
while the first integral (over negative values of $F$) is dominated by the 
saddle-point contribution of left tail of the function $P(F)$, namely
\begin{equation}
\label{22}
\int_{-\infty}^{0} dF \, \exp\bigl\{-\gamma \, |F|^{\alpha} \, + \, \beta N |F|\bigr\}
\; \sim \; 
\exp\Bigl\{g \, (\beta N)^{k} \, t\Bigr\}
\end{equation}
Formally the asymptotic form, eq.(\ref{20}), is valid only for large (negative)
values of $F$ so that the integration here can not be extended up to zero.
Nevertheless, since the saddle-point value of the free energy,
\begin{equation}
\label{23}
|F_{*}| \; = \; \biggl(\frac{\beta N}{\alpha \, \gamma}\biggr)^{\frac{1}{\alpha - 1}}
\end{equation}
is big  in the limit $N \gg 1$, the contribution of the integration over 
finite (negative) values of $F$ (where PDF $P(F)$ is not known) can be neglected
compared to the saddle-point contribution in the asymptotic region $F\to -\infty$.
Thus, substituting this saddle-point value into the exponential in 
the l.h.s of eq.(\ref{22}) we get
\begin{equation}
\label{24}
\exp\Bigl\{ \frac{\alpha}{\alpha-1}\bigl(\alpha\gamma\bigr)^{-\frac{1}{\alpha-1}} \,
           \bigl(\beta N\bigr)^\frac{\alpha}{\alpha-1}   \Bigr\}
 \; \sim \; 
\exp\Bigl\{g \, (\beta N)^{k} \, t\Bigr\}
\end{equation}
From this relation we immediately obtain
\begin{equation}
\label{25}
\alpha \; = \; \frac{k}{k-1}
\end{equation}
and 
\begin{equation}
\label{26}
\gamma \; = \;  \frac{k-1}{k} \, \bigl(k \, g\, t\bigr)^{-\frac{1}{k-1}}
\end{equation}
Substituting these values of $\alpha$ and $\gamma$ into eq.(\ref{20}) 
we get the following {\it left} asymptotics of the free energy probability distribution
function
\begin{equation}
\label{27}
P(F\to -\infty) \; \sim \; 
\exp\biggl\{
-\frac{k-1}{k} \biggl(\frac{|F|}{\bigl(k\, g \, t\bigr)^{1/k}}\biggr)^{\frac{k}{k-1}} 
\biggr\}
\end{equation}
Now, if we admit that the free energy fluctuations of the considered system
are described by a {\it universal} probability distribution function, then
we can conclude that the time scaling of these fluctuations can be represented as
\begin{equation}
\label{28}
F \; \sim \; f \cdot t^{1/k}
\end{equation}
where the random quantity  $f \sim 1$ is described by a probability  distribution function
${\cal P}(f)$ whose left tail is 
\begin{equation}
\label{29}
{\cal P}(f\to -\infty) \; \sim \; \exp\Bigl\{ - \mbox{(const)} \, |f|^{k/(k-1)}\Bigr\}
\end{equation} 
while its entire form remains unknown.

\section{Replica energy balance method}

In view of the absence of the exact solution for the $N$-particle 
two-dimensional wave function $\psi({\bf r}_{1},  \, ... \, {\bf r}_{N})$ 
we propose the following 
"heuristic" strategy of calculations of the replica energy $E_{N}$ defined by eq.(\ref{13})
in the limit $N \gg 1$. Let us multiply eq.(\ref{13}) by 
$\psi({\bf r}_{1},  \, ... \, {\bf r}_{N}) \; \delta \bigl(\sum_{a=1}^{N}{\bf r}_{a}\bigr)$
and integrate over all particle's positions:
\begin{eqnarray}
\nonumber
&&-2 \beta \, E_{N} \, \int \, d^{2} r_{1} ... d^{2} r_{N}
\delta \Bigl(\sum_{a=1}^{N}{\bf r}_{a}\Bigr)
\psi^{2}({\bf r}_{1},  \, ... \, {\bf r}_{N}) 
= \sum_{a=1}^{N} \, 
\int \, d^{2} r_{1} ... d^{2} r_{N} \; 
\delta \Bigl(\sum_{a=1}^{N}{\bf r}_{a}\Bigr) 
\psi({\bf r}_{1},  \, ... \, {\bf r}_{N})\Delta_{a} 
\psi({\bf r}_{1},  \, ... \, {\bf r}_{N}) 
\\
\nonumber
\\
&&
+  \kappa \, \sum_{a\not= b}^{N} 
\int \, d^{2} r_{1} ... d^{2} r_{N}  
\delta \Bigl(\sum_{a=1}^{N}{\bf r}_{a}\Bigr) 
\delta^{(2d)}_{\epsilon}({\bf r}_{a} - {\bf r}_{b}) \,
\psi^{2}({\bf r}_{1},  \, ... \, {\bf r}_{N})
 +  
\frac{\kappa}{\pi\epsilon^{2}} \, N 
\int \, d^{2} r_{1} ... d^{2} r_{N} 
\delta \Bigl(\sum_{a=1}^{N}{\bf r}_{a}\Bigr) 
\psi^{2}({\bf r}_{1},  \, ... \, {\bf r}_{N})
\label{30}
\end{eqnarray}
Taking into account the symmetry of the wave function with respect to the particles
permutations the above equation can be rewritten as follows
\begin{eqnarray}
\nonumber
&&2 \beta \, \tilde{E}_{N} \, \int \, d^{2} r_{1} ... d^{2} r_{N} \; 
\delta \Bigl(\sum_{a=1}^{N}{\bf r}_{a}\Bigr)
\psi^{2}({\bf r}_{1},  \, ... \, {\bf r}_{N}) 
= N \, 
\int \, d^{2} r_{1} ... d^{2} r_{N} \; 
\delta \Bigl(\sum_{a=1}^{N}{\bf r}_{a}\Bigr) 
\Bigl(\nabla_{1}\psi({\bf r}_{1},  \, ... \, {\bf r}_{N})\Bigr)^{2}
\\
\nonumber
\\
&&
- \;   \kappa \,N (N-1) 
\int \, d^{2} r_{1} ... d^{2} r_{N}  \;
\delta \Bigl(\sum_{a=1}^{N}{\bf r}_{a}\Bigr) 
\delta^{(2d)}_{\epsilon}({\bf r}_{1} - {\bf r}_{2}) \,
\psi^{2}({\bf r}_{1},  \, ... \, {\bf r}_{N})
\label{31}
\end{eqnarray}
where
\begin{equation}
\label{32}
\tilde{E}_{N} \; = \; E_{N} \; + \; \frac{\kappa N}{2\beta \, \epsilon^{2}}
\end{equation}

Let us suppose that the wave function of the present system is characterized by the only
spatial scale $R$ (which will be defined later by minimizing the energy $\tilde{E}_{N}$),
$\psi({\bf r}_{1},  \, ... \, {\bf r}_{N}) \; \to \; 
\psi({\bf r}_{1}/R,  \, ... \, {\bf r}_{N}/R)$. Taking into account the properties of 
wave function of the two-particle system, eqs.(\ref{15}), where
the Macdonald function
\begin{equation}
\label{33}
K_{0}(x) \; \sim \;  
\left\{ 
\begin{array}{ll}
\ln(1/x)  
\; ,  \; \; \; \; \; \; \; 
\mbox{at} \; x \ll 1 
\\
\\                          
\exp\{-x\}  \; , \; \; \; \;  
\mbox{at} \; x \gtrsim 1 
\end{array}
\right.
\end{equation}
we will suppose that the $N$-particle wave function 
$\psi({\bf r}_{1}/R,  \, ... \, {\bf r}_{N}/R)$
exhibits two principal properties: 

(1) for any two particles $a$ and $b$, such that $ \epsilon < |{\bf r}_{a} - {\bf r}_{b}|  \ll R$,
\begin{equation}
\label{34}
\psi\bigl({\bf r}_{1}/R, \, ..., \, 
{\bf r}_{a}/R, \, ..., \, {\bf r}_{b}/R, \, ..., \, {\bf r}_{N}/R \bigr) \; \sim \;  
\ln\Bigl(\frac{R}{|{\bf r}_{a} - {\bf r}_{b}|}\Bigr) \times
\psi\bigl({\bf r}_{1}/R, \, ..., \, 
{\bf r}_{b}/R, \, ..., \, {\bf r}_{b}/R, \, ..., \, {\bf r}_{N}/R \bigr)
\end{equation}

(2) at scales $|{\bf r}_{a} - {\bf r}_{b}|  \gtrsim R$,
\begin{equation}
\label{35}
\psi\bigl({\bf r}_{1}/R, \, ..., \,  {\bf r}_{N}/R \bigr) \; \sim \;  
\exp\Bigl\{ 
- \frac{1}{R} \sum_{1\leq a < b}^{N} \, |{\bf r}_{a} - {\bf r}_{b}|
\Bigr\}
\end{equation}
After rescaling ${\bf r}_{a} \to R \, {\bf x}_{a}$ the above wave function 
$\psi\bigl({\bf r}_{1}/R, \, ..., \,  {\bf r}_{N}/R \bigr) \; \to \; 
\psi({\bf x}_{1},  \, ... \, {\bf x}_{N})$, and at scales $|{\bf x}_{a} - {\bf x}_{b}|  \gtrsim 1$
instead of eq.(\ref{35}) we get
\begin{equation}
\label{35a}
\psi({\bf x}_{1},  \, ... \, {\bf x}_{N}) \; \sim \;  
\exp\Bigl\{ 
-  \sum_{1\leq a < b}^{N} \, |{\bf x}_{a} - {\bf x}_{b}|
\Bigr\}
\end{equation}
Now the relation (\ref{31}) takes the form
\begin{equation}
\label{36}
2\beta \tilde{E}_{N}(R) \; I_{0}(N) \; = \; 
N \, R^{-2} \, I_{\small{\nabla}}(N, R) \; - \; 
N (N-1) \, \kappa \, R^{-2} \, I_{\delta} (N,R)
\end{equation}
where
\begin{equation}
\label{37}
I_{0}(N) \; = \;  \int \, d^{2} x_{1} ... d^{2} x_{N} \; 
\delta \Bigl(\sum_{a=1}^{N}{\bf x}_{a}\Bigr)
\psi^{2}({\bf x}_{1},  \, ... \, {\bf x}_{N}) 
\end{equation}
\begin{equation}
\label{38}
I_{\nabla}(N,R) \; = \;  \int \, d^{2} x_{1} ... d^{2} x_{N} \; 
\delta \Bigl(\sum_{a=1}^{N}{\bf x}_{a}\Bigr)
\Bigl(\nabla_{1}\psi({\bf x}_{1},  \, ... \, {\bf x}_{N})\Bigr)^{2}
\end{equation}
\begin{equation}
\label{39}
I_{\delta}(N,R) \; = \;  \int \, d^{2} x_{1} ... d^{2} x_{N} \; 
\delta \Bigl(\sum_{a=1}^{N}{\bf x}_{a}\Bigr)
\delta^{(2d)}_{\epsilon/R}({\bf x}_{1} - {\bf x}_{2}) \,
\psi^{2}({\bf x}_{1},  \, ... \, {\bf x}_{N}) 
\end{equation}
According to eq.(\ref{34}), the leading (most divergent) 
contribution to the function $I_{\nabla}(N,R)$, eq.(\ref{38}),
comes from the integrations over ${\bf x}_{1}$ in the vicinities of every point 
${\bf x}_{a}$ ($a = 2, ...N$):
\begin{equation}
\label{40}
I_{\nabla}(N,R) \; \sim \;  \int \, d^{2} x_{2} ... d^{2} x_{N} \; 
\sum_{a=2}^{N} \; \int_{\frac{\epsilon}{R} < |{\bf x}_{1} - {\bf x}_{a}| \ll 1}
d^{2} x_{1} \Bigl[\nabla_{1} \ln\Bigl(\frac{1}{|{\bf x}_{1} - {\bf x}_{a}|}\Bigr)\Bigr]^{2}
\delta \Bigl({\bf x_{a}} + \sum_{b=2}^{N}{\bf x}_{b}\Bigr)
\psi^{2}\bigl({\bf x}_{a}, \, {\bf x}_{2}, \, ..., \, 
{\bf x}_{a}, \, ... , \, {\bf x}_{N} \bigr)
\end{equation}
Using the symmetry of the wave function $\psi({\bf x}_{1},  \, ... \, {\bf x}_{N})$
and redefining ${\bf x}_{1} = {\bf x}_{a} + {\boldsymbol \eta}$
we get
\begin{equation}
\label{41}
I_{\nabla}(N,R) \; \sim \; (N-1) \, \int \, d^{2} x_{2} ... d^{2} x_{N} \; 
\int_{\epsilon/R}^{1} d\eta \, \eta \, \Bigl[\frac{d}{d\eta} \ln\Bigl(\frac{1}{\eta}\Bigr)\Bigr]^{2}
\delta \Bigl({\bf x_{2}} + \sum_{b=2}^{N}{\bf x}_{b}\Bigr)
\psi^{2}\bigl({\bf x}_{2}, \, {\bf x}_{2}, \, {\bf x}_{3}, \, ..., \, {\bf x}_{N} \bigr)
\end{equation}
or
\begin{equation}
\label{42}
I_{\nabla}(N,R) \; \sim \; (N-1) \, \ln\Bigl(\frac{R}{\epsilon}\Bigr)
\int \, d^{2} x_{2} ... d^{2} x_{N} \; 
\delta \Bigl({\bf x_{2}} + \sum_{b=2}^{N}{\bf x}_{b}\Bigr) \, 
\psi^{2}\bigl({\bf x}_{2}, \, {\bf x}_{2}, \, {\bf x}_{3}, \, ..., \, {\bf x}_{N} \bigr)
\end{equation}
In a similar way we can estimate the value of the function $I_{\delta}(N,R)$, eq.(\ref{39}).
Taking into account that according to eq.(\ref{34})
\begin{equation}
\label{43}
\lim_{|{\bf x}_{1} - {\bf x}_{2}|\to \epsilon/R} \; 
\psi^{2}\bigl({\bf x}_{1}, \, {\bf x}_{2}, \, {\bf x}_{3}, \, ..., \, {\bf x}_{N} \bigr)
\; \sim \;
 \ln^{2}\Bigl(\frac{R}{\epsilon}\Bigr) \; 
\psi^{2}\bigl({\bf x}_{2}, \, {\bf x}_{2}, \, {\bf x}_{3}, \, ..., \, {\bf x}_{N} \bigr) 
\end{equation}
and that $\int d^{2}x \, \delta^{(2d)}_{\epsilon/R}({\bf x}) \, = \, 1$, we obtain
\begin{equation}
\label{44}
I_{\delta}(N,R) \; \sim \;  \ln^{2}\Bigl(\frac{R}{\epsilon}\Bigr)
\int \, d^{2} x_{2} ... d^{2} x_{N} \; 
\delta \Bigl({\bf x_{2}} + \sum_{b=2}^{N}{\bf x}_{b}\Bigr) \, 
\psi^{2}\bigl({\bf x}_{2}, \, {\bf x}_{2}, \, {\bf x}_{3}, \, ..., \, {\bf x}_{N} \bigr)
\end{equation}
Substituting eqs.(\ref{42}) and (\ref{44}) into eq.(\ref{36}) we find
\begin{equation}
\label{45}
2\beta \tilde{E}_{N}(R) \; I_{0}(N) \; \sim \; 
N(N-1) \, \Bigl[R^{-2} \, \ln\Bigl(\frac{R}{\epsilon}\Bigr) \; - \; 
 \kappa \, R^{-2} \,  \ln^{2}\Bigl(\frac{R}{\epsilon}\Bigr) \Bigr] \; 
 I_{1}(N) 
\end{equation}
or
\begin{equation}
\label{46}
\tilde{E}_{N}(R)  \; \sim \; \frac{N(N-1)}{2\beta}
 \, \Bigl[R^{-2} \, \ln\Bigl(\frac{R}{\epsilon}\Bigr) \; - \; 
\kappa \, R^{-2} \,  \ln^{2}\Bigl(\frac{R}{\epsilon}\Bigr) \Bigr] \; 
\frac{I_{1}(N)}{I_{0}(N)} 
\end{equation}
where
\begin{equation}
\label{47}
I_{1}(N) \; = \;  \int \, d^{2} x_{2} ... d^{2} x_{N} \; 
\delta \Bigl({\bf x_{2}} + \sum_{b=2}^{N}{\bf x}_{b}\Bigr) \, 
\psi^{2}\bigl({\bf x}_{2}, \, {\bf x}_{2}, \, {\bf x}_{3}, \, ..., \, {\bf x}_{N} \bigr)
\end{equation}
and $I_{0}(N)$ is given in eq.(\ref{37}).

Further strategy consist in two (independent) steps: we have to find optimal 
parameter $R_{*}$ which minimizes $\tilde{E}_{N}(R)$, and we have to calculate
the ratio $I_{1}(N)/I_{0}(N)$ as a function of $N$ 
(in the limit $N\to \infty$). In this way we will find the 
energy $E_{*}(N) \equiv \tilde{E}_{N}(R_{*})$
of considered $N$-particle system as a function of $N$ which defines
the replica partition function $Z(N,t) \sim \exp\{-t \, E_{*}(N)\}$.
This would be sufficient (as it was shown in Section II) to get the scaling exponent
of the free energy fluctuations as well as the form of the left tail of its probability
distribution function. But first, let us demonstrate how the proposed method
works in the case of $(1+1)$ system where we can compare obtained results with 
well known exact ones.

\section{(1+1) directed polymers}

In one dimension instead of eq.(\ref{13}) we have
the Schr\"odinger equation for the $N$-particle wave function
$\psi(r_{1}, \, ... \, r_{N})$ which depends on $N$ one-dimensional 
particles coordinates
$\{-\infty \, < \, r_{a} \, < \, +\infty\} \; (a = 1, ..., N)$:
\begin{equation}
\label{48}
-2 \beta \, E_{N} \, 
\psi(r_{1}, \, ... \, r_{N}) \; = \;
\sum_{a=1}^{N} \, \frac{\partial^{2}}{\partial r_{a}^{2}} \,  
\psi(r_{1}, \, ... \, r_{N})
\; + \; \kappa \, \sum_{a,b=1}^{N} \delta_{\epsilon}(r_{a} - r_{b}) \,
\psi(r_{1}, \, ... \, r_{N})
\end{equation}
where, as before, $\kappa = \beta^{3} u$ and $\delta_{\epsilon}(r)$ is 
the "finite-size $\delta$-function"
\begin{equation}
\label{49}
\delta_{\epsilon}(r) \; = \; \left\{ 
\begin{array}{ll}
\frac{1}{2\epsilon} 
\; ,  
\; \; \; \;
\mbox{for} \; |r| \, \leq \, \epsilon 
\\
\\                          
0 \; , 
\; \; \; \; \; \;  
\mbox{for} \;  |r| \, > \, \epsilon                 
\end{array}
\right.
\end{equation}
Multiplying eq.(\ref{48})  by 
$\psi(r_{1}, \, ... \, r_{N}) \; \delta \bigl(\sum_{a=1}^{N} r_{a}\bigr)$
and integrating over all particle's positions instead of eq.(\ref{30}) 
we get:
\begin{eqnarray}
\nonumber
&&2 \beta \, \tilde{E}_{N} \, \int_{-\infty}^{+\infty} \, dr_{1} ... dr_{N} \, 
\delta \Bigl(\sum_{a=1}^{N} \, r_{a}\Bigr) \, 
\psi^{2}(r_{1}, \, ... \, r_{N})
= \sum_{a=1}^{N} \, 
\int_{-\infty}^{+\infty} \, dr_{1} ... dr_{N} \, 
\delta \Bigl(\sum_{a=1}^{N} \, r_{a}\Bigr) \, 
\Bigl(\frac{\partial}{\partial r_{a}} \,\psi(r_{1},  \, ... \, r_{N})\Bigr)^{2}
\\
\nonumber
\\
&&
-  \kappa \, \sum_{a\not= b}^{N} 
\int_{-\infty}^{+\infty} \, dr_{1} ... dr_{N} \, 
\delta \Bigl(\sum_{a=1}^{N} \, r_{a}\Bigr) \, 
\delta_{\epsilon}(r_{a} - r_{b}) \,
\psi^{2}(r_{1},  \, ... \, r_{N})
\label{50}
\end{eqnarray}
where
\begin{equation}
\label{51}
\tilde{E}_{N} \; = \; E_{N} \; + \; \frac{\kappa N}{4\beta \, \epsilon}
\end{equation}

Supposing, as before, that the wave function of the present system is characterized 
by the only spatial scale $R$ and taking into account the properties of the 
two-particle solution we admit that the $N$-particle wave function has the
Bethe anzats structure,
\begin{equation}
\label{52}
\psi(r_{1},  \, ... \, r_{N}) \; = \;  
\exp\Bigl\{ 
- \frac{1}{R} \sum_{1\leq a < b}^{N} \, |r_{a} - r_{b}|
\Bigr\}
\end{equation}

After rescaling $r_{a} \to R \, x_{a}$ the relation (\ref{50}) takes the form
\begin{equation}
\label{53}
2\beta \tilde{E}_{N}(R) \; I_{0}(N) \; = \; 
 R^{-2} \, I_{\nabla}(N) \; - \; 
 \kappa \, R^{-1} \, I_{\delta} (N)
\end{equation}
where
\begin{equation}
\label{54}
I_{0}(N) \; = \;  \int \, dx_{1} ... dx_{N} \; 
\delta \Bigl(\sum_{a=1}^{N} x_{a}\Bigr) \,
\psi^{2}(x_{1},  \, ... \, x_{N}) 
\end{equation}
\begin{equation}
\label{55}
I_{\nabla}(N) \; = \;  \int \, dx_{1} ... dx_{N} \; 
\delta \Bigl(\sum_{a=1}^{N} x_{a}\Bigr) \,
\sum_{b=1}^{N} \,
\Bigl(\frac{\partial}{\partial x_{a}}
\psi(x_{1},  \, ... \, x_{N})\Bigr)^{2}
\end{equation}
\begin{equation}
\label{56}
I_{\delta}(N) \; = \;  \int \, dx_{1} ... dx_{N} \; 
\delta \Bigl(\sum_{a=1}^{N} x_{a}\Bigr) \,
\sum_{b\not= c}^{N} \,
\delta (x_{b} - x_{c}) \,
\psi^{2}(x_{1},  \, ... \, x_{N}) 
\end{equation}
and 
\begin{equation}
\label{57}
\psi(x_{1},  \, ... \, x_{N}) \; = \;  
\exp\Bigl\{ 
-\sum_{1\leq a < b}^{N} \, |x_{a} - x_{b}|
\Bigr\}
\end{equation}
As the wave function $\psi(x_{1},  \, ... \, x_{N})$, eq.(\ref{57}), exhibits no 
divergences at $ |x_{a} - x_{b}| \to 0$,
the expression for $I_{\delta}(N)$, eq.(\ref{56}), is given in the limit $\epsilon \to 0$.

Detailed calculations of the above three factors are given in Appendix A.
The results are:
\begin{eqnarray}
\label{58}
I_{0}(N) &=& 2^{-(N-1)} \, \frac{1}{(N-1)!}
\\
\nonumber
\\
\label{59}
I_{\nabla}(N) &=& \frac{1}{3} \, N (N^{2} - 1) \; I_{0}(N) 
\\
\nonumber
\\
\label{60}
I_{\delta}(N) &=& \frac{1}{3} \, N (N^{2} - 1) \; I_{0}(N)
\end{eqnarray}
Substituting these expressions into eq.(\ref{53}) we get
 \begin{equation}
 \label{61}
 \tilde{E}_{N}(R)  \; = \; 
 \frac{1}{6\beta} \,  N (N^{2} - 1) \, 
 \bigl( R^{-2} \; - \; \kappa \, R^{-1} \bigr)
  \end{equation}
The minimum of $\tilde{E}_{N}(R)$ is achieved at $R_{*} \, = \, 2/\kappa$
such that 
\begin{equation}
\label{62}
\tilde{E}_{N}(R_{*}) \; \equiv \; E_{*}(N) \; = \; 
-\frac{\kappa^{2}}{24\beta} \, (N^{3} \, - \, N)
\end{equation}
Substituting here $\kappa \, = \, \beta^{3} \, u$ we get
\begin{equation}
\label{63}
E_{*}(N) \; = \; 
-\frac{1}{24} \, \beta^{5} \, u^{2} \, (N^{3} \, - \, N)
\end{equation}
which perfectly coincides with the exact result for the ground state 
energy of $N$-particle boson system with paired attractive $\delta$ interactions
(see e.g. \cite{VD-TW-2010-2}). In the limit $N \to \infty$ we obtain
\begin{equation}
\label{64}
Z(N,t) \; \sim \; \exp\bigl\{- E_{*}(N) \, t\bigr\} 
\; \sim \; \exp\Bigl\{ \frac{1}{24} (\beta u)^{2} \, (\beta N)^{3}  \Bigr\}
\end{equation}
so that according to the discussion in Section II, eqs.(\ref{19}) and 
(\ref{27})-(\ref{29}), the free energy fluctuations scale with time as
\begin{equation}
\label{65}
F \; \sim \;  (\beta u)^{2/3} \, t^{1/3} \cdot f
\end{equation}
where the random quantity $f \sim 1$ is described by a probability distribution function
with the left tail
\begin{equation}
\label{66}
{\cal P}(f\to -\infty) \; \sim \; 
\exp\Bigl\{ - (\mbox{const}) \, |f|^{3/2}\Bigr\}
\end{equation}
which also perfectly fits with well known exact results 
(see e.g. \cite{LeDoussal-2010,VD-TW-2010-1}).

\section{(2+1) Directed polymers: free energy scaling}

According to eq.(\ref{46}), to derive the scaling exponent of the free energy
fluctuations of (2+1) directed polymers we have to find the minimum of $\tilde{E}_{N}(R)$
with respect to $R$ (which is easy), but more important, we hate to derive the asymptotics
of the ratio $I_{1}(N)/I_{0}(N)$ in the limit $N\to\infty$ (which is not easy).
We start with the treatment of the factor $I_{1}(N)$, eq.(\ref{47}):
\begin{equation}
\label{67}
I_{1}(N) \; = \;  \int \, d^{2} x_{2} ... d^{2} x_{N} \; 
\delta \Bigl({\bf x_{2}} + \sum_{b=2}^{N}{\bf x}_{b}\Bigr) \, 
\exp\Bigl\{ 
-2\sum_{a=3}^{N} |{\bf x}_{2} - {\bf x}_{a}|  \; - \; 
2\sum_{2\leq a<b}^{N} |{\bf x}_{a} - {\bf x}_{b}| 
\Bigr\}
\end{equation}
Here and in what follows we approximate the wave function 
$\psi\bigl({\bf x}_{1}, \, ..., \, {\bf x}_{N} \bigr)$ 
by its form at scales $|{\bf x}_{a} - {\bf x}_{b}|  \gtrsim 1$, eq.(\ref{35a}).
Note that as the integration in the vicinities of the logarithmic singularities at 
$|x_{a}-x_{b}| \to \epsilon \ll 1$ is
converging its contribution is negligible compared to the one at scales
$|{\bf x}_{a} - {\bf x}_{b}|  \gtrsim 1$.

It can be easily shown (see Appendix B) that at $N \gg 1$ both
the typical (average) value of the distances between the particles
$\langle |{\bf x}_{a} - {\bf x}_{b}| \rangle \; \sim \; 1/N$
and the typical value of the lengths of the vectors 
$\langle |{\bf x}_{a}| \rangle \; \sim \; 1/N$.
It makes possible to compare the orders of magnitudes 
of the two factors in the exponential in the r.h.s. of eq.(\ref{67}):
\begin{eqnarray}
\nonumber
&&
\sum_{a=3}^{N} |{\bf x}_{2} - {\bf x}_{a}| \; \sim \; 
\frac{1}{N} \, (N-2) \; \sim \; 1
\\
\label{68}
\\
\nonumber
&&
\sum_{2\leq a<b}^{N} |{\bf x}_{a} - {\bf x}_{b}| \; \sim \; 
\frac{1}{N} \, (N-1)(N-2) \; \sim \; N
\end{eqnarray}
which shows that we can neglect the first one compared to the second one.
Similarly, for the terms under the $\delta$-function  of eq.(\ref{67})
we have:
\begin{eqnarray}
\nonumber
&&
{\bf x}_{2} \; \sim \; \frac{1}{N} 
\\
\label{69}
\\
\nonumber
&&
\sum_{b=2}^{N} {\bf x}_{a}  \; \sim \; 
\frac{1}{N} \, (N-1) \; \sim \; 1
\end{eqnarray}
which makes possible to neglect the first term compared to the second one.

Thus, we conclude that
\begin{equation}
\label{70}
I_{1}(N) \; \sim \; 
\int \, d^{2} x_{2} ... d^{2} x_{N} \; 
\delta \Bigl(\sum_{b=2}^{N}{\bf x}_{b}\Bigr) \, 
\exp\Bigl\{ -2\sum_{2\leq a<b}^{N} |{\bf x}_{a} - {\bf x}_{b}| 
\Bigr\} \; = \; I_{0}(N-1)
\end{equation}
The factor $I_{0}(N)$, eq.(\ref{37}), can be represented as follows:
\begin{eqnarray}
\nonumber
I_{0}(N) &\sim& 
\int \, d^{2}x_{1} \, d^{2}x_{2} \,  ... \, d^{2}x_{N} \; 
\delta \Bigl(\sum_{a=1}^{N}{\bf x}_{a}\Bigr) \, 
\exp\Bigl\{ -2\sum_{1\leq a<b}^{N} |{\bf x}_{a} - {\bf x}_{b}| 
\Bigr\}
\\
\nonumber
\\
&=&
\int d^{2}x_{2} ... d^{2}x_{N} \; \int d^{2}x_{1} \; 
\delta \Bigl({\bf x}_{1} + \sum_{a=2}^{N}{\bf x}_{a}\Bigr) \, 
\exp\Bigl\{ 
-2\sum_{a=2}^{N} |{\bf x}_{1} - {\bf x}_{a}|  \; - \; 
2\sum_{2\leq a<b}^{N} |{\bf x}_{a} - {\bf x}_{b}| 
\Bigr\}
\label{71}
\end{eqnarray}
At this stage we are going to integrate over one degree of freedom (over ${\bf x}_{1}$)
which would make possible to relate the factors $I_{0}(N)$ and $I_{0}(N-1)$. 

Let us redefine the integration variables: 
$\{{\bf x}_{2}, \, ... \, , {\bf x}_{N}\} \; \to \; 
\{ {\bf z}, \, {\bf y}_{2}, \, ... \, , {\bf y}_{N-1} \}$:
\begin{eqnarray}
\nonumber
{\bf x}_{a} &=& {\bf z} \; + \; {\bf y}_{a} \; \; \; \; \;  (a \, = \, 2, ..., N-1)
\\
\label{72}
{\bf x}_{N} &=& {\bf z} \; - \; \sum_{a=2}^{N-1} \, {\bf y}_{a}
\end{eqnarray}
Denoting ${\bf y}_{N} \equiv -\sum_{a=2}^{N-1} \, {\bf y}_{a}$, 
instead of eq.(\ref{71}) we get
\begin{equation}
\label{73}
I_{0}(N) \; \sim \; 
\int d^{2}z \int d^{2}y_{2} ... d^{2}y_{N} \, 
\delta\Bigl(\sum_{a=2}^{N} {\bf y}_{a} \Bigr) \; 
D(N) \; 
\int d^{2}x_{1} \, \delta\bigl({\bf x}_{1} \, + \, (N-1){\bf z}\bigr)
\exp\Bigl\{ 
-2\sum_{a=2}^{N} |{\bf x}_{1} - {\bf z} - {\bf y}_{a}|  \; - \; 
2\sum_{2\leq a<b}^{N} |{\bf y}_{a} - {\bf y}_{b}| 
\Bigr\}
\end{equation}
where $D(N)$ is the Jacobian
\begin{equation}
\label{74}
D(N) \; = \; \det\bigg|\frac{\partial\bigl({\bf x}_{2} , ..., {\bf x}_{N}\bigr)}{
               \partial\bigl({\bf z}, {\bf y}_{2}, ..., {\bf y}_{N-1}\bigr)} \bigg|
\end{equation}
Simple calculations yield (see Appendix C)
\begin{equation}
\label{75}
D(N) \; = \; (N-1)^{2} \, \Big|_{N\gg 1}  \; \simeq \; N^{2}
\end{equation}
Thus, integrating over ${\bf x}_{1}$ in eq.(\ref{73}), we get
\begin{equation}
\label{76}
I_{0}(N) \; \sim \; N^{2} 
\int d^{2}y_{2} ... d^{2}y_{N} \; 
\delta\Bigl(\sum_{a=2}^{N} {\bf y}_{a} \Bigr) \; 
\exp\Bigl\{
-2\sum_{2\leq a<b}^{N} |{\bf y}_{a} - {\bf y}_{b}| 
\Bigr\} \,
\int d^{2}z \; 
\exp\Bigl\{ -2\sum_{a=2}^{N} |N {\bf z} + {\bf y}_{a}| \Bigr\}
\end{equation}
The above expression can be represented as follows:
\begin{equation}
\label{77}
I_{0}(N) \; \sim \; N^{2} \, I_{0}(N-1) \,
\int d^{2}z \, 
\frac{\int d^{2}y_{2} ... d^{2}y_{N} \; 
	\delta\Bigl(\sum_{a=2}^{N} {\bf y}_{a} \Bigr) \; 
	\exp\Bigl\{
	-2\sum_{2\leq a<b}^{N} |{\bf y}_{a} - {\bf y}_{b}| 
	\Bigr\} \, \exp\Bigl\{ -2\sum_{a=2}^{N} |N {\bf z} + {\bf y}_{a}| \Bigr\}}{
\int d^{2}y_{2} ... d^{2}y_{N} \; 
\delta\Bigl(\sum_{a=2}^{N} {\bf y}_{a} \Bigr) \; 
\exp\Bigl\{
-2\sum_{2\leq a<b}^{N} |{\bf y}_{a} - {\bf y}_{b}| 
\Bigr\} }
\end{equation}
or
\begin{equation}
\label{78}
I_{0}(N) \; \sim \; N^{2} \, I_{0}(N-1) \,
\int d^{2}z \;
\bigg\langle \exp\Bigl\{ -2\sum_{a=2}^{N} |N {\bf z} + {\bf y}_{a}| \Bigr\}
\bigg\rangle_{\{{\bf y}\}}
\end{equation}
where the averaging is done over $\{{\bf y}_{2}, ..., {\bf y_{N}}\}$
with the wight 
$\exp\bigl\{ -2\sum_{2\leq a<b}^{N} |{\bf y}_{a} - {\bf y}_{b}| \bigr\} $.
Redefining ${\bf z} = \frac{1}{2N^{2}} {\bf u}$ and 
${\bf y}_{a} = \frac{1}{2N} {\bf v}_{a}$ we get
\begin{equation}
\label{79}
I_{0}(N) \; \sim \; \frac{1}{4N^{2}} \, I_{0}(N-1) \,
\int d^{2}u \;
\bigg\langle \exp\Bigl\{ -\frac{1}{N} \sum_{a=2}^{N} |{\bf u} + {\bf v}_{a}| \Bigr\}
\bigg\rangle_{\{{\bf v}\}}
\end{equation}
where the averaging is done over $\{{\bf v}_{2}, ..., {\bf v_{N}}\}$
with the wight 
$\exp\bigl\{ -\frac{1}{N}\sum_{2\leq a<b}^{N} |{\bf v}_{a} - {\bf v}_{b}| \bigr\} $.
As the typical value of ${\bf y}_{a}$ 
(defined by the distribution 
$\exp\bigl\{ -2\sum_{2\leq a<b}^{N} |{\bf y}_{a} - {\bf y}_{b}| \bigr\} $) 
is $\sim N^{-1}$ (see Appendix B), the  typical value of the vectors ${\bf v}_{a}$ 
is $\sim 1$. Thus, we can estimate
\begin{equation}
\label{80}
\Phi({\bf u}) \equiv
\bigg\langle \exp\Bigl\{ -\frac{1}{N} \sum_{a=2}^{N} |{\bf u} + {\bf v}_{a}| \Bigr\}
\bigg\rangle_{\{{\bf v}\}}
\; \sim \; 
 \left\{ 
\begin{array}{ll}
\bigg\langle \exp\Bigl\{ -\frac{1}{N} \sum_{a=2}^{N} |{\bf v}_{a}| \Bigr\}
\bigg\rangle_{\{{\bf v}\}} \, = \, C_{0}, \sim 1
\; ,  
\; \; \; \;
\mbox{at} \; |{\bf u}| \, \ll 1 
\\
\\                          
\bigg\langle \exp\Bigl\{ -\frac{1}{N} \sum_{a=2}^{N} |{\bf u}| \Bigr\}
\bigg\rangle_{\{{\bf v}\}} \, \simeq   \exp\bigl\{ -|{\bf u}| \bigr\} , 
\; \; \; \; \; \; \; 
\mbox{at} \;  |{\bf u}| \, \gg 1                
\end{array}
\right.
\end{equation}
which means, according to eq.(\ref{79}), that
\begin{equation}
\label{81}
I_{0}(N) \; \sim \; \frac{1}{4N^{2}} \, I_{0}(N-1) \,
\int d^{2}u \;
\Phi({\bf u}) \; \ \sim \; N^{-2} \, I_{0}(N-1)
\end{equation}
or, according to eq.(\ref{70})
\begin{equation}
\label{82}
I_{0}(N-1) \; \sim \; I_{1}(N) \; \sim \;  N^{2} \, I_{0}(N)
\end{equation}
Substituting this into eq.(\ref{46}), we get
\begin{equation}
\label{83}
\tilde{E}_{N}(R)  \; \sim \; \frac{1}{\beta} \, N^{4} \, 
\, \Bigl[R^{-2} \, \ln\Bigl(\frac{R}{\epsilon}\Bigr) \; - \; 
\kappa \, R^{-2} \,  \ln^{2}\Bigl(\frac{R}{\epsilon}\Bigr) \Bigr] 
\end{equation}
This function of $R$ has the minimum at
\begin{equation}
\label{84}
\ln\Bigl(\frac{R_{*}}{\epsilon}\Bigr) \; = \; 
\frac{1}{2\kappa} \, 
\bigl(1 \, + \, \kappa \, + \, \sqrt{1 \, + \, \kappa^{2}}\bigr) \, \Big|_{\kappa \gg 1}
\; \simeq \; 1
\end{equation}
and its value at the minimum
\begin{equation}
\label{85}
\tilde{E}_{N}(R_{*}) \, \equiv \, E_{*}(N) \, \sim \, 
- \frac{\kappa}{\beta \epsilon^{2}} \, N^{4} \; = \; 
-\frac{u}{({\beta \, \epsilon})^{2}} \, \bigl(\beta \, N \bigr)^{4} 
\end{equation}
Correspondingly, for the replica partition function we get
\begin{equation}
\label{86}
Z(N, t) \; \sim \; \exp\bigl\{- E_{*}(N) \, t \bigr\} \; \sim \; 
\exp\Bigl\{
\frac{u}{({\beta \, \epsilon})^{2}} \, \bigl(\beta \, N \bigr)^{4} \, t
\Bigr\}
\end{equation}
Thus, according to eqs.(\ref{19}) and (\ref{27})-(\ref{29}), the free energy fluctuations
in the considered system scale with time as
\begin{equation}
\label{87}
F \; \sim \; \Bigl(\frac{\sqrt{u}}{\beta \, \epsilon}\Bigr)^{1/2} \, t^{1/4} \cdot f
\end{equation}
where the random quantity $f \sim 1$ is described by a probability distribution function
with the left tail
\begin{equation}
\label{88}
{\cal P}(f\to -\infty) \; \sim \; 
\exp\Bigl\{ - (\mbox{const}) \, |f|^{4/3}\Bigr\}
\end{equation}
The above eqs.(\ref{87}) and (\ref{88}) constitute the main result of the present research.

\section{Discussion}

In this paper we argue that at least in the low temperature limit the scaling exponent 
of the free energy fluctuations of $(2+1)$ directed polymers is equal to $\theta=1/4$ which
close but not equal to the value $0.241$ obtained in numerical simulations. 
I would suggest that this discrepancy could be due to the presence of 
a logarithmic prefactor which in numerical simulations may distort the "pure" 
scaling behavior with $\theta = 1/4$ and which is out of reach of the present 
analytical method.

It is worthing to stress that according to eq.(\ref{87}), the prefactor in the time scaling of
the free energy fluctuation explicitly depend on the size $\epsilon$ of the spatial
correlations of the random potential, eq.(\ref{4}). Moreover, we see that the limit 
$\epsilon \to 0$ (when these correlations turn into the true $\delta$-function) 
do not exist here (the prefactor is divergent). 
In other words, one can not consider the (2+1) directed polymers
with $\delta$-correlated random potential because such system is pathological.  
This is in drastic difference with (1+1) directed polymers where it is just for the 
$\delta$-correlated random potential that the exact solution has been derived.

Of course, present research is based on the calculations which are far from being rigorous,
and correspondingly it leaves many open questions. The main problem is that the ground state 
wave function for the considered two-dimensional $N$-particle boson system with 
short-range attractive interactions (which should be the solution of eq.(\ref{13})) 
is not known. The key assumption of the present approach is that this wave function
can be approximated by the heuristic structure given in eqs.(\ref{34})-(\ref{35}),
which is a straightforward generalization of the two-particle solution for the 
$N$-particle system and which contains one (spatial scale) adjustment parameter.
It is evident that such structure of the wave function is not correct for an arbitrary
(finite) value of $N$, but on the other hand it seems to be a good approximation 
in the limit $N \to \infty$. To what extend the use of such approximation is essential 
for the declared result $\theta =1/4$ is not clear.

Another key point of the present approach is the {\it hypothesis} that the free energy
fluctuations of the considered system are described by a {\it universal} 
probability distribution function. This is known to take place in (1+1) directed polymers
where the exact solution is available, but it is not necessary guaranteed for the considered system. 
If this hypothesis is correct then for obtaining the value of the scaling exponent
it is sufficient to derive the left tail of the free energy distribution function 
(that is what has been done in this paper). If not, then the knowledge  of this left tail
tells us just nothing.

One more problem is the evident dependence of the scaling properties of the considered system
on the temperature. The present study was devoted to the low temperature limit when
dimensionless parameter $\kappa = \beta^{3} \, u \; \gg \; 1$ (see eqs.(\ref{84})-(\ref{85}))
and in this case we get $\theta = 1/4$.  
Unfortunately in the framework of the present approach one can not study the high temperature
limit (when $\kappa = \beta^{3} \, u \; \ll \; 1$) as in this case the minimum of the function 
$\tilde{E}_{N}(R)$, eq.(\ref{83}), becomes almost "flat" and correspondingly the 
value of $R_{*}$ becomes ill defined. On the other hand, previous analytical study
of the high temperature limit (using completely different method) \cite{(2+1)-highT}
yields the result $\theta = 1/2$. If these two results  
are both correct then we would have to conclude that unlike (1+1) directed polymers the value of the
scaling exponent in the present system is non-universal being temperature dependent.
 
Finally, like in the case of (1+1) directed polymers the {\it zero temperature} limit 
($\beta \to \infty$) in the considered (2+1) system remains non accessible in the framework 
of the usual continuous-space and {\it replica symmetric} approach. 
Either we have to consider the lattice version of this model or we have to introduce 
some kind of a replica symmetry breaking ansatz \cite{zero-T}. 

In any case further systematic study (both theoretical an numerical) 
of this system would be highly welcomed.

\newpage


\vspace{10mm}

\begin{center}
	
	\appendix{\large \bf Appendix A}

\end{center}

\newcounter{A}
\setcounter{equation}{0}
\renewcommand{\theequation}{A.\arabic{equation}}

\vspace{5mm}

For the calculations of the integrals in eqs.(\ref{54})-(\ref{56})
due to the symmetry of the wave function $\psi(x_{1},  \, ... \, x_{N})$  
we can order the particle's positions, e.g. as follows:
\begin{equation}
\label{A1}
\{ x_{1}, x_{2}, ..., x_{N} \} \; \to \; 
\{ -\infty < x_{1} \leq x_{2} \leq ... \leq x_{N} < +\infty \}
\end{equation}
In this case
\begin{equation}
\label{A2}
\int \, dx_{1} \, ... \, dx_{N} \; \to \; 
N!  
\underbrace{\int dx_{1} \, ... \, dx_{N}}_{-\infty < x_{1} \leq ... \leq x_{N} < +\infty} 
\end{equation}
and 
\begin{equation}
\label{A3}
\sum_{1\leq a < b}^{N} \, |x_{a} - x_{b}| \; = \; 
-\sum_{k=1}^{N} \bigl(N - 2k +1 \bigr) \, x_{k}
\end{equation}
so that the wave function takes the form
\begin{equation}
\label{A4}
\psi(x_{1},  \, ... \, x_{N}) \; = \; 
\exp\Bigl\{  \sum_{k=1}^{N} \, \mu_{k} \, x_{k}  \Bigr\}
\end{equation}
where 
\begin{equation}
\label{A5}
\mu_{k} \; = \; \bigl(N - 2k +1 \bigr)
\end{equation}

For the factor $I_{0}(N)$, eq.(\ref{54}), we get the following integral representation
\begin{equation}
\label{A6}
I_{0}(N) \; = \; 2 \,
N! \, \int_{-\infty}^{+\infty} \frac{dp}{2\pi} 
\int_{-\infty}^{+\infty}dx_{N} \int_{-\infty}^{x_{N}} dx_{N-1} \; ... \;
\int_{-\infty}^{x_{3}} dx_{2} \int_{-\infty}^{x_{2}} dx_{1}
\exp\Bigl\{2\sum_{k=1}^{N} \bigl(\mu_{k} + ip\bigr) x_{k} \Bigr\}
\end{equation}
Integrating one after another over $x_{1}, x_{2}, ..., x_{N-1}$ we obtain
\begin{equation}
\label{A7}
I_{0}(N) \; = \; 2^{-(N-1)} \, N!  \,
\int_{-\infty}^{+\infty} \frac{dp}{2\pi} 
\int_{-\infty}^{+\infty}dx_{N}
\frac{\exp\Bigl\{ \Bigl(\sum_{k=1}^{N} \mu_{k} \; + \; iNp\Bigr) \, x_{N} \Bigr\}}{
(\mu_{1} + ip)(\mu_{1} + \mu_{2} + 2ip) \, ... \, (\mu_{1} + ... + \mu_{N-1} + i(N-1)p)}
\end{equation}
Taking into account that $\sum_{k=1}^{N} \mu_{k} \, = \, 0$ we eventually find
\begin{eqnarray}
\nonumber
I_{0}(N) &=& 2^{-(N-1)} \, N!  \,
\int_{-\infty}^{+\infty} dp 
\frac{\delta(N p)}{
(\mu_{1} + ip)(\mu_{1} + \mu_{2} + 2ip) \, ... \, (\mu_{1} + ... + \mu_{N-1} + i(N-1)p)}
\\
\nonumber
\\
\nonumber
&=&
2^{-(N-1)} \, N!  \, \frac{1}{N} \, 
\frac{1}{\mu_{1}(\mu_{1} + \mu_{2}) \, ... \, (\mu_{1} + ... + \mu_{N-1})}
\\
\nonumber
\\
\nonumber
&=&
2^{-(N-1)} \, (N - 1)!  \,
\frac{1}{(N-1)(2N-4) (3N-9) \, ... \, \bigl[(N-1) - (N-1)(N-1) \bigr] }
\\
\nonumber
\\
\nonumber
&=&
2^{-(N-1)} \, (N - 1)!  \, \frac{1}{\bigl[(N-1)!\bigr]^{2}}
\\
\nonumber
\\
&=&
\frac{2^{-(N-1)}}{(N-1)!}
\label{A8}
\end{eqnarray}

In a similar way for the factor $I_{\nabla}(N)$, eq.(\ref{55}),
we obtain
\begin{eqnarray}
\nonumber
I_{\nabla}(N) &=& 
2 \, N! \, \sum_{k=1}^{N} \mu_{k}^{2} \;
\int_{-\infty}^{+\infty} \frac{dp}{2\pi} 
\int_{-\infty}^{+\infty}dx_{N} \int_{-\infty}^{x_{N}} dx_{N-1} \; ... \;
\int_{-\infty}^{x_{2}} dx_{1} \;
\exp\Bigl\{2\sum_{k=1}^{N} \bigl(\mu_{k} + ip\bigr) x_{k} \Bigr\}
\\
\nonumber
\\
\nonumber
&=&
\sum_{k=1}^{N} (N - 2k +1)^{2} \; I_{0}(N)
\\
\nonumber
\\
&=&
\frac{1}{3} \, N (N^{2}-1) \; I_{0}(N)
\label{A9}
\end{eqnarray}

The calculations of $I_{\delta}(N)$, eq.(\ref{56}), are slightly  more cumbersome:
\begin{eqnarray}
\nonumber
I_{\delta}(N) &=&
N! \; \sum_{k=1}^{N-1}  
\underbrace{\int dx_{1} \, ... \, dx_{N}}_{-\infty < x_{1} \leq ... \leq x_{N} < +\infty} 
\delta\bigl(x_{k}-x_{k+1}\bigr) \, \delta\Bigl(\sum_{l=1}^{N} x_{l}\Bigr) \,
\exp\Bigl\{2 \sum_{n=1}^{N} \mu_{n} \, x_{n} \Bigr\}
\\
\nonumber
\\
\nonumber
&=&
2^{-(N-2)} \, N! \, \int_{-\infty}^{+\infty} \frac{dp}{2\pi}
 \sum_{k=1}^{N-1}  
\underbrace{\int dx_{1} \, ... \, dx_{k} \, \int dx_{k+2} \, ... \, dx_{N}}_{-\infty < x_{1} \leq ... \leq x_{k} \leq x_{k+2} \leq ... \leq x_{N} < +\infty}
\; \times 
\\
\nonumber
\\
&\times&
\exp\Bigl\{ \sum_{n=1}^{k-1} \bigl(\mu_{n}+ip\bigr) \, x_{n} 
\, + \, 
\bigl(\mu_{k} + \mu_{k+1} + 2ip\bigr) \, x_{k} 
\, + \,
\sum_{n=k+2}^{N} \bigl(\mu_{n}+ip\bigr) \, x_{n} 
\Bigr\}
\label{A10}
\end{eqnarray}
Integrating over $\{x_{1}, ..., x_{k}, x_{k+2}, ..., x_{N} \}$ 
(s.f. eqs.(\ref{A6})-(\ref{A8}))
and taking into account that $\sum_{n=1}^{N} \mu_{n} = 0$, we get
\begin{eqnarray}
\nonumber
I_{\delta}(N) &=&
2^{-(N-2)} \, N! \, \int_{-\infty}^{+\infty} dp \sum_{k=1}^{N-1} \; 
\frac{ \delta\bigl( N \, p\bigr)}{\mu_{1} \, (\mu_{1} + \mu_{2}) \, ... \, 
(\mu_{1} + ... + \mu_{k-1}) \times (\mu_{1} + ... + \mu_{k+1}) \, ... \,
(\mu_{1} + ... + \mu_{N-1}) }
\\
\nonumber
\\
\nonumber
&=&
2^{-(N-2)} \, (N-1)! \sum_{k=1}^{N-1} \;
\frac{ (\mu_{1} + \mu_{2} + ... + \mu_{k})}{
\mu_{1} \, (\mu_{1} + \mu_{2}) \, ... \, (\mu_{1} + ... + \mu_{N-1})}
\\
\nonumber
\\
\nonumber
&=&
2^{-(N-2)} \, (N-1)! \, \frac{1}{\bigl( (N-1)!\bigr)^{2}} \, 
\sum_{k=1}^{N-1} \sum_{n=1}^{k} \, (N -2n + 1)
\\
\nonumber
\\
\nonumber
&=&
2 \, I_{0}(N) \, \sum_{k=1}^{N-1} \, (N - k) \, (N - 2k +1)
\\
\nonumber
\\
&=&
\frac{1}{3}\,  N (N^{2} - 1) \, I_{0}(N)
\label{A11}
\end{eqnarray}


\vspace{10mm}

\begin{center}
	
	\appendix{\large \bf Appendix B}

\end{center}

\newcounter{B}
\setcounter{equation}{0}
\renewcommand{\theequation}{B.\arabic{equation}}

\vspace{5mm}

The average distance between particles can be computed as follows
\begin{equation}
\label{B1}
\langle |\Delta {\bf x}| \rangle \; = \; 
\frac{2}{N(N-1)} \sum_{a<b}^{N} \langle |{\bf x}_{a} - {\bf x}_{b}| \rangle \; = \;
\frac{2\int  d^{2}x_{1} \,  ... \, d^{2}x_{N} \; 
	\delta \Bigl(\sum_{a=1}^{N}{\bf x}_{a}\Bigr) \,
	\sum_{a<b}^{N} |{\bf x}_{a} - {\bf x}_{b}| \; 
\exp\Bigl\{ -2\sum_{1\leq a<b}^{N} |{\bf x}_{a} - {\bf x}_{b}| 
\Bigr\} 
}{N(N-1) \; 
\int  d^{2}x_{1} \,  ... \, d^{2}x_{N} \; 
\delta \Bigl(\sum_{a=1}^{N}{\bf x}_{a}\Bigr) \,
\exp\Bigl\{ -2\sum_{1\leq a<b}^{N} |{\bf x}_{a} - {\bf x}_{b}| \Bigr\}
}
\end{equation}
or
\begin{equation}
\label{B.2}
\langle |\Delta {\bf x}| \rangle \; = \; 
-\frac{2}{N(N-1)} \; \frac{\partial}{\partial \lambda} \; 
\ln\biggl[
\int  d^{2}x_{1} \,  ... \, d^{2}x_{N} \; 
\delta \Bigl(\sum_{a=1}^{N}{\bf x}_{a}\Bigr) \,
\exp\Bigl\{ -\lambda \sum_{1\leq a<b}^{N} |{\bf x}_{a} - {\bf x}_{b}| \Bigr\}
\biggr] \, \bigg|_{\lambda =2}
\end{equation}
After rescaling ${\bf x}_{a} \to \lambda^{-1} \, {\bf x}_{a}$ we get 
\begin{eqnarray}
\nonumber
\langle |\Delta {\bf x}| \rangle &=& 
-\frac{2}{N(N-1)} \; \frac{\partial}{\partial \lambda} \; 
\ln\biggl[
\lambda^{-2(N-1)}
\int  d^{2}x_{1} \,  ... \, d^{2}x_{N} \; 
\delta \Bigl(\sum_{a=1}^{N}{\bf x}_{a}\Bigr) \,
\exp\Bigl\{ -\sum_{1\leq a<b}^{N} |{\bf x}_{a} - {\bf x}_{b}| \Bigr\}
\biggr] \, \bigg|_{\lambda =2}
\\
\nonumber
\\
\nonumber
&=&
-\frac{2}{N(N-1)} \; \frac{\partial}{\partial \lambda} \; 
\biggl\{
-2(N-1) \ln (\lambda) \; + \; 
\ln\biggl[
\int  d^{2}x_{1} \,  ... \, d^{2}x_{N} \; 
\delta \Bigl(\sum_{a=1}^{N}{\bf x}_{a}\Bigr) \,
\exp\Bigl\{ -\sum_{1\leq a<b}^{N} |{\bf x}_{a} - {\bf x}_{b}| \Bigr\}
\biggr]
\biggr\} \, \bigg|_{\lambda =2}
\\
\nonumber
\\
\nonumber
&=&
\frac{2}{N(N-1)} \, 2(N-1) \, \frac{1}{\lambda} \, \Big|_{\lambda =2} \; = \; \frac{2}{N}
\\
\nonumber
\\
\label{B.3}
&\sim&
\, \frac{1}{N}
\end{eqnarray}
As for any two particles $ |{\bf x}_{a} - {\bf x}_{b}| \, \sim \, N^{-1}$,
the square of this difference $ \bigl({\bf x}_{a} - {\bf x}_{b}\bigr)^{2} \, \sim \, N^{-2}$,
or
\begin{equation}
\label{B.4}
\frac{1}{N(N-1)} \sum_{a,b}^{N} \bigl({\bf x}_{a} - {\bf x}_{b}\bigr)^{2} 
\; = \; 
\frac{1}{N(N-1)} \Bigl[2N \sum_{a=1}^{N} {\bf x}_{a}^{2} \; - \; 
2 \, \Bigl(\sum_{a=1}^{N} {\bf x}_{a} \Bigr)^{2} \Bigr]
\; \sim \; N^{-2}
\end{equation}
Taking into account that $\sum_{a=1}^{N} {\bf x}_{a} \; = \; 0$ we obtain
\begin{equation}
\label{B.5}
\frac{2}{(N-1)} \,\sum_{a=1}^{N} {\bf x}_{a}^{2} \; \sim \; N^{-2}
\end{equation} 
which means that ${\bf x}_{a}^{2} \, \sim \, N^{-2}$ and $|{\bf x}_{a}| \, \sim \, N^{-1}$.


\vspace{10mm}

\begin{center}
	
	\appendix{\large \bf Appendix C}

\end{center}

\newcounter{C}
\setcounter{equation}{0}
\renewcommand{\theequation}{C.\arabic{equation}}

\vspace{5mm}

The Jacobian of the linear transformation
\begin{eqnarray}
\nonumber
{\bf x}_{a} &=& {\bf z} \; + \; {\bf y}_{a} \; \; \; \; \;  (a \, = \, 2, ..., N-1)
\\
\label{C.1}
{\bf x}_{N} &=& {\bf z} \; - \; \sum_{a=2}^{N-1} \, {\bf y}_{a}
\end{eqnarray}
is
\begin{equation}
\label{C.2}
D(N) \; = \; \det\bigg|\frac{\partial\bigl({\bf x}_{2} , ..., {\bf x}_{N}\bigr)}{
	\partial\bigl({\bf z}, {\bf y}_{2}, ..., {\bf y}_{N-1}\bigr)} \bigg|
\end{equation}
In the Cartesian vector representation ${\bf x}_{a} = (p_{a}, q_{a}) \, ; \; \; 
{\bf y}_{a} = (s_{a}, t_{a}) \, ; \; \; {\bf z}_{a} = (u, v)$
the above linear transformation reads
\begin{eqnarray}
\nonumber
p_{a} &=& u \; + \; s_{a} \; \; \; \; \;  (a \, = \, 2, ..., N-1)
\\
\label{C.3}
p_{N} &=& u \; - \; \sum_{a=2}^{N-1} \, s_{a}
\\
\nonumber
\\
\nonumber
q_{a} &=& v \; + \; t_{a} \; \; \; \; \;  (a \, = \, 2, ..., N-1)
\\
\label{C.4}
q_{N} &=& v \; - \; \sum_{a=2}^{N-1} \, t_{a}
\end{eqnarray}
so that
\begin{equation}
\label{C.5}
D(N) \; = \; \det\bigg|\frac{\partial\bigl(p_{2} , ..., p_{N}\bigr)}{
	\partial\bigl(u, s_{2}, ..., s_{N-1}\bigr)} \bigg| \, \times \, 
\det\bigg|\frac{\partial\bigl(q_{2} , ..., q_{N}\bigr)}{
	\partial\bigl(v, t_{2}, ..., t_{N-1}\bigr)} \bigg|
\end{equation}
Explicitly,
\begin{equation}
\label{C.6}
\det\bigg|\frac{\partial\bigl(p_{2} , ..., p_{N}\bigr)}{
	\partial\bigl(u, s_{2}, ..., s_{N-1}\bigr)} \bigg| \; = \; \det \, 
\begin{tabular}{|c|c|c|c|c|}
\hline
$\frac{\partial p_{2}}{\partial u}$  & $\frac{\partial p_{3}}{\partial u}$ & $\frac{\partial p_{4}}{\partial u}$ & ... & $\frac{\partial p_{N}}{\partial u}$ \\
\hline
$\frac{\partial p_{2}}{\partial s_{2}}$ & $\frac{\partial p_{3}}{\partial s_{2}}$ & $\frac{\partial p_{4}}{\partial s_{2}}$ & \; ... \; & $\frac{\partial p_{N}}{\partial s_{2}}$\\ 
\hline
$\frac{\partial p_{2}}{\partial s_{3}}$ & $\frac{\partial p_{3}}{\partial s_{3}}$  &$\frac{\partial p_{4}}{\partial s_{3}}$   & ... & $\frac{\partial p_{N}}{\partial s_{3}}$ \\ 
\hline
\multicolumn{5}{|c|}{. . . . . . . . . . . . }  \\ 
\hline
$\frac{\partial p_{2}}{\partial s_{N-1}}$ & $\frac{\partial p_{3}}{\partial s_{N-1}}$ & $\frac{\partial p_{4}}{\partial s_{N-1}}$ &   ... & $\frac{\partial p_{N}}{\partial s_{N-1}}$\\ 
\hline
\end{tabular}
\end{equation}
Simple calculations yield
\begin{equation}
\label{C.7}
\det\bigg|\frac{\partial\bigl(p_{2} , ..., p_{N}\bigr)}{
	\partial\bigl(u, s_{2}, ..., s_{N-1}\bigr)} \bigg| \; = \; \det \, 
\begin{tabular}{|c|c|c|c|c|}
\hline
$\; \; 1 \; \; $  & $\; \; 1 \; \; $ & $\; \; 1 \; \; $ & ... & $1$ \\
\hline
$1$ & $0$ & $0$ & \; ... \; & $-1$\\ 
\hline
$0$ & $1$  &$0$   & ... & $-1$ \\ 
\hline
\multicolumn{5}{|c|}{. . . . . . . . . . . . }  \\ 
\hline
$0$ & $0$ & $0$ &   ... & $-1$\\ 
\hline
\end{tabular}
\; = \; 
(-1)^{N} \, (N-1)
\end{equation}
Evidently, one gets the same result for the second determinant in the r.h.s of eq.(\ref{C.5}).
Thus, we obtain
\begin{equation}
\label{C.8}
D(N) \; = \; (N-1)^{2} \Big|_{N\gg 1} \; \simeq \; N^{2}
\end{equation}
%


\end{document}